\documentclass[10pt]{iopart}
\usepackage{epsfig}
\usepackage{psfrag}
\usepackage{iopams}

\newcommand{\ket}[1]{| #1 \rangle}
\newcommand{\bra}[1]{\langle #1 |}

\newcommand{\bdr}{{\bf r}}

\psfrag{xlabela}{$10^{-3}N_0$}
\psfrag{xlabelb}{$10^{-3}N_0$}
\psfrag{ylabela}{$\Omega_c$ (units of $\omega$)}
\psfrag{ylabelb}{$\Omega_c$ (units of $\omega$)}
\psfrag{fig1Axlabel}{$r$ (units of $\ell_{\rm trap}$)}
\psfrag{fig1Bxlabel}{$r$ (units of $\ell_{\rm trap}$)}
\psfrag{fig1Aylabel}{$\psi(\bdr)$}
\psfrag{fig1Bylabel}{${2m \over \hbar^2} g_{\rm 2D}(\bdr)$}
\psfrag{fig2Axlabel}{$\mu$ (units of $\hbar\omega$)}
\psfrag{fig2Aylabel}{$10^{-4}N_0$}
\psfrag{fig2Bxlabel}{$\mu$ (units of $\hbar\omega$)}
\psfrag{fig2Bylabel}{Percentage difference in $N_0$}

\begin{document}

\title{Solutions of the Gross-Pitaevskii equation in two dimensions}
\author{M~D~Lee\dag and S~A~Morgan\ddag} 
\address{\dag\ Clarendon Laboratory, Department of Physics, University of Oxford,
Parks Road, Oxford OX1~3PU, U.K.} 
\address{\ddag\ Department of Physics and Astronomy, University College London, 
Gower Street, London WC1E~6BT, U.K.}
\date{\today}

\begin{abstract} 
In two dimensions the Gross-Pitaevskii equation for a cold, dilute gas of bosons
has an energy dependent coupling parameter describing particle interactions.  We
present numerical solutions of this equation for bosons in harmonic traps and
show that the results can be quite sensitive to the precise form of the coupling
parameter that is used.
\end{abstract} 

\maketitle

\section{Introduction}

The experimental realization of Bose-Einstein condensation (BEC) in dilute atomic
gases has prompted a great deal of work on the theoretical description of such
systems.  Underlying many theoretical treatments is the notion that the condensate
behaviour is governed, to first order, by the Gross-Pitaevskii equation (GPE). 
Recent experiments~\cite{gorlitz2001} and 
proposals~\cite{zobay2001,gauck1998,hinds1998} for BEC in two dimensional (2D)
traps have triggered interest in the properties of condensates in 2D.  Although
long wavelength fluctuations prohibit BEC in a 2D homogeneous system at any finite
temperature~\cite{hohenberg1967}, the presence of a trapping potential alters the
density of states sufficiently that finite temperature 2D Bose condensates may be
created in trapped systems~\cite{bagnato1991}.  Such condensates obey a modified
form of the Gross-Pitaevskii equation, which contains the correct description of
scattering in 2D.  In a recent paper~\cite{lee2002} we derived this description
based on an approximation of the many-body T-matrix in terms of the off-shell
two-body T-matrix.  In this paper we present numerical solutions of the 2D GPE for
ground state and vortex states of a zero temperature gas of bosons in a harmonic
trap, and contrast the results with other forms that have been suggested
recently~\cite{kolomeisky2000}.  We show that the detailed numerical predictions
are quite sensitive to the precise form of the interaction strength used.  

In the following section we summarize the scattering theory needed to describe
interactions in a 2D homogeneous BEC, before discussing in
section~\ref{sec:inhomog} the various approximations by which these results may
be applied to a trapped gas.  Finally, in section~\ref{sec:results} we present
numerical results for each of these approximations and discuss the significant
differences which can arise between them.

\section{The Gross-Pitaevskii equation and scattering}

The condensate wave function $\psi(\bdr)$ is given by the solution of the two
dimensional Gross-Pitaevskii equation
\begin{equation}
-{\hbar^2 \over 2m} \nabla^2 \psi(\bdr) + V_{\rm trap}(\bdr) \psi(\bdr) +
N_0 
g_{\rm 2D}(\bdr)\left|
\psi(\bdr)\right|^2 \psi(\bdr) = \mu \psi(\bdr), \label{eq:GPE}
\end{equation}
where $V_{\rm trap}(\bdr)$ is an external trapping potential (which is generally
harmonic in present BEC experiments), $N_0$ is the condensate population, and
$\mu$ is the chemical potential.  The coupling parameter $g_{\rm 2D}(\bdr)$
appearing in the non-linear term describes the interactions between two
condensate atoms.

A collision between two atoms in momentum states $\ket{k}$ and $\ket{m}$ which
produces a transition to states $\ket{i}$ and $\ket{j}$ is described by the
T-matrix element $\bra{ij}T(E)\ket{km}$, where $E$ is the energy of the
collision. The T-matrix is obtained as the solution of a Lippmann-Schwinger
equation or equivalently via a summation of ladder diagrams~\cite{taylor}.  In
three-dimensional (3D) systems the coupling parameter in the GPE is often taken to
be the zero-energy, zero-momentum limit of the two-body T-matrix, which
describes the scattering of particles in a vacuum. This gives $\bra{00}T_{\rm
2b}(0)\ket{00} =g_{\rm 3D} = 4\pi\hbar^2a_{\rm 3D}/m$ where $a_{\rm 3D}$ is the
\emph{s}-wave scattering length~\cite{taylor}.  This is, however, merely an
approximate description since the scattering of two condensate particles
actually occurs in a medium consisting of the surrounding particles rather than
in a vacuum.  Instead the collision is properly described by a many-body
T-matrix element $\bra{00}T_{\rm MB}(0)\ket{00}$ which incorporates the effects
of the surrounding atoms on the scattering process. In 3D the
many-body T-matrix leads to a relatively small correction to the two-body
T-matrix approximation (of relative order $(na_{\rm 3D}^3)^{1/2}$ at $T=0$) and
for many purposes it is sufficient to neglect many-body effects in the GPE.  In
two dimensions (and lower), however, the two-body T-matrix vanishes in the
zero-energy, zero-momentum limit~\cite{morgan2002}, and many-body effects are
therefore of much greater importance and contribute even at leading order.

In a recent paper~\cite{lee2002} we demonstrated how the many-body T-matrix can
be approximated by the two-body T-matrix evaluated off the energy shell. The
coupling constant which appears in the GPE in a homogeneous 2D system was shown
to be~\cite{lee2002}
\begin{equation}
g_{\rm 2D} = \bra{00}T_{\rm MB}(0)\ket{00} =
\bra{00}T_{\rm 2b}(-\mu)\ket{00}.
\end{equation}
Using the expression for the off-shell two-body T-matrix found
in reference~\cite{morgan2002}, gives the following form of the coupling
parameter in a homogeneous system~\cite{lee2002}
\begin{equation}
g_{\rm 2D} =  
-{4\pi \hbar^2 \over m}{1 \over \ln\left(\mu ma_{\rm 2D}^2/4\hbar^2\right)},
\label{eq:g2d}
\end{equation}
where  $a_{\rm 2D}$ is a two-dimensional scattering length.   For a 2D gas of
hard spheres of radius $a$ we have $a_{\rm 2D} = ae^{\gamma_{\rm EM}}$, where
$\gamma_{\rm EM} \approx 0.577$ is the Euler-Mascheroni constant which we have
absorbed into the definition of $a_{\rm 2D}$ here for convenience\footnote{Note that this definition of $a_{\rm 2D}$ differs sightly from
that in our earlier work such that our $a_{\rm 2D}$ here equals $a_{\rm
2D}e^{\gamma_{\rm EM}}$ in the notation of reference~\cite{morgan2002}.  The
definition used here simplifies the form of equation~(\ref{eq:g2d}).}. 
In practice, a 2D gas is created by trapping atoms very tightly in one
dimension (the $z$ axis) such that the motion in the $z$ direction is
effectively frozen out. The effective 2D scattering length for such a gas is
given by~\cite{petrov2000,petrov2001}
\begin{equation} 
a_{\rm
2D} = 4\sqrt{\pi \over B}l_z \exp\left(-\sqrt{\pi}{l_z \over a_{\rm 3D}}\right),
\label{eq:a2d}
\end{equation} 
where $B \approx 0.915$, $a_{\rm 3D}$ is the 3D \emph{s}-wave scattering
length, and $l_z = \sqrt{\hbar/2m\omega_z}$ is the typical width of the system
in the $z$ direction.  The 2D scattering length therefore depends
not only on the 3D scattering length, but also upon the degree of confinement
in the $z$ direction.  

Equation~(\ref{eq:g2d}) shows that the coupling parameter for a 2D homogeneous
Bose gas depends on the chemical potential of the system (and hence the
density) as well as the 2D scattering length.  This is in contrast with the
case in 3D where the coupling parameter depends only upon the
scattering length to first order.

\section{The GPE in trapped 2D systems}
\label{sec:inhomog}

The expression for the 2D coupling parameter shown in equation~(\ref{eq:g2d})
was derived for a homogeneous system, and the correct application of these
results to the case of a trapped gas is the main objective of this paper.  In a
previous paper~\cite{lee2002} we provided solutions of the 2D GPE in a trap
using the homogeneous coupling parameter of equation~(\ref{eq:g2d}) in
order to illustrate the effect of the energy dependence of $g_{\rm 2D}$.  In
this paper we focus on a more accurate description of the scattering in
inhomogeneous systems.

We consider a 3D Bose gas confined tightly in one dimension and weakly in the
remaining two dimensions on a length scale $\ell_{\rm trap}$.  A collision between
two condensate particles will typically occur over some length scale $\ell_{\rm
coll}$.  Provided that $\ell_{\rm coll}$ is much smaller than $\ell_{\rm trap}$ we
can use a local density approximation.  

We can introduce the length scale $\ell_{\rm coll}$ by the following simple
argument.  We model the pair wave function of two atoms in the medium by that of a
single particle with the reduced mass moving in a potential which consists of a
circularly symmetric box of radius $L$ and a hard sphere of radius $a$ located in
the centre of the box.  For \emph{s}-wave scattering the wave function is solved by
\begin{equation}
\psi(r) =  A_0J_0(kr) + B_0 N_0(kr),
\end{equation}
where $J_0$ and $N_0$ are Bessel functions of the first and second kind
respectively.  In the zero-energy, zero-momentum limit we get $\psi(r) = A_0 +
B_0 \ln (r)$.  Applying the boundary conditions that the wave function vanishes
on the radius $r=a$ and reaches the asymptotic value $\chi$ at the edge of the
box gives
\begin{equation}
\psi(r) = \chi{\ln (r/a) \over \ln (L/a)}  \mbox{   for } a < r < L.
\end{equation}
The extra energy caused by the curvature of this wave function
resulting from the presence of the scattering potential is
\begin{equation}
\Delta E = {\hbar^2 \over 2m} \int_a^L |\nabla\psi(r)|^2 d^2r =
{\hbar^2 |\chi|^2 \over 2m} {2\pi \over \ln (L/a)}. \label{eq:delE}
\end{equation}

This energy depends upon the size of the box $L$, which is indeed the length scale
relevant for the scattering of two particles in 2D.  The scattering of two
particles in a many-body system should obviously not depend on the size of the
system as a whole when $L$ becomes large, and so we must interpret $L$ as the
physically relevant length scale $\ell_{\rm coll}$.  The appropriate length scale
over which a many-body wave function changes is the healing length $\ell_h$, given
in homogeneous Bose condensed systems by $\ell_h =  \hbar/\sqrt{2mg_{\rm 2D}n_0} =
\hbar/\sqrt{2m\mu}$, and so it is this which must be used in
equation~(\ref{eq:delE}).  Since $N_0|\chi|^2$ corresponds to the condensate
density $n_0$, this leads in the homogeneous limit to a pair interaction strength
of the form of equation~(\ref{eq:g2d}).

The same argument can be applied straightforwardly to trapped gases if the
condensate density varies slowly on the scale of the healing length, which is true
except in the surface region.  In this case, many-body effects cause the pair wave
function to reach its asymptotic value on a length scale equal to the local healing
length $\ell_h = \hbar/\sqrt{2mg_{\rm 2D}(\bdr)n_0(\bdr)}$.  The two-body
interaction strength is therefore given from equation~(\ref{eq:g2d}) by replacing
$\mu$ with $n_0(\bdr)g_{\rm 2D}(\bdr)$ producing a density-dependent effective
interaction.  Such density dependent coupling parameters are expected from the
results of density functional theory~\cite{nunes1999} which predict that the energy
of the system is a functional of the density only.

In an inhomogeneous system the density is spatially dependent and thus the
coupling parameter is also spatially dependent.  In terms of the condensate
wave function the density is given by $n_0(\bdr) = N_0|\psi(\bdr)|^2$. 
Equation~(\ref{eq:g2d}) now gives for the coupling parameter the result
\begin{equation}
g_{\rm 2D}(\bdr) = 
-{4\pi \hbar^2 \over m}{1 \over \ln\left(N_0|\psi(\bdr)|^2 g_{\rm 2D}(\bdr) 
ma_{\rm 2D}^2/4\hbar^2\right)}.
\label{eq:gn0}
\end{equation}
An approximate solution to this equation may be found by iteration, giving
\begin{equation}
g_{\rm 2D}(\bdr) = -{4\pi\hbar^2 \over m} \left[ \ln(N_0\pi |\psi(\bdr)|^2a_{\rm
2d}^2)
\right]^{-1} +O\left(\ln[\ln(n_0a_{\rm 2D}^2)]\over \ln(n_0a_{\rm 2D}^2)\right). 
\label{eq:gn0simple}
\end{equation}
The first order term in this expansion agrees with the form of coupling parameter
proposed by Kolomeisky \emph{et  al.\ }~\cite{kolomeisky2000,kolomeisky1992} who
used the renormalization group to analyse a 2D homogeneous Bose gas.  Earlier work
by Shevchenko~\cite{shevchenko1992} also proposed such a coupling parameter based
on the work of Schick~\cite{schick}.  More recently Tanatar \emph{et al.\
}\cite{tanatar2002} have also made use of this approximation. Unfortunately the
expansion in equation~(\ref{eq:gn0simple}) is not valid for realistic systems since
the higher order terms are not in general negligible, as will be shown in the
following section.

A more accurate procedure would be to solve equation~(\ref{eq:gn0}) numerically
for $g_{\rm 2D}(n_0(\bdr))$ and use this exact solution in solving the GPE.  In the
following section we present results which suggest that this accurate solution
may be necessary in some circumstances.

\section{Solutions to the Gross-Pitaevskii Equation}
\label{sec:results}

In this section we present numerical solutions of the 2D GPE for a Bose condensate
trapped in a circularly symmetric, harmonic potential characterized by an angular
frequency $\omega$.  The various approximation schemes for the coupling parameter
discussed in the previous section will be compared to the more accurate $g_{\rm
2D}(\bdr)$ found from the numerical solution of equation~(\ref{eq:gn0}).  In order
to illustrate our results, we choose as our parameters $\omega = 2\pi \times 100
{\rm Hz}$ and $a_{\rm 2D} = 6 {\rm nm}$.  This 2D scattering length is
approximately the 3D scattering length found for $^{87}{\rm Rb}$, and so from
equation~(\ref{eq:a2d}) this corresponds to a situation where $l_z \approx a_{\rm
3D}$, and hence the two-dimensional nature of the scattering is
important~\cite{lee2002}.  For these parameters the results we obtain correspond
to  healing lengths between $0.1 \ell_{\rm trap}$ and $0.5 \ell_{\rm trap}$ (for
the very low $\mu$ solutions) at the centre of the trap.  These healing lengths are
sufficiently small for the local density approximation to be valid.

\begin{figure}
\center{
\epsfig{file=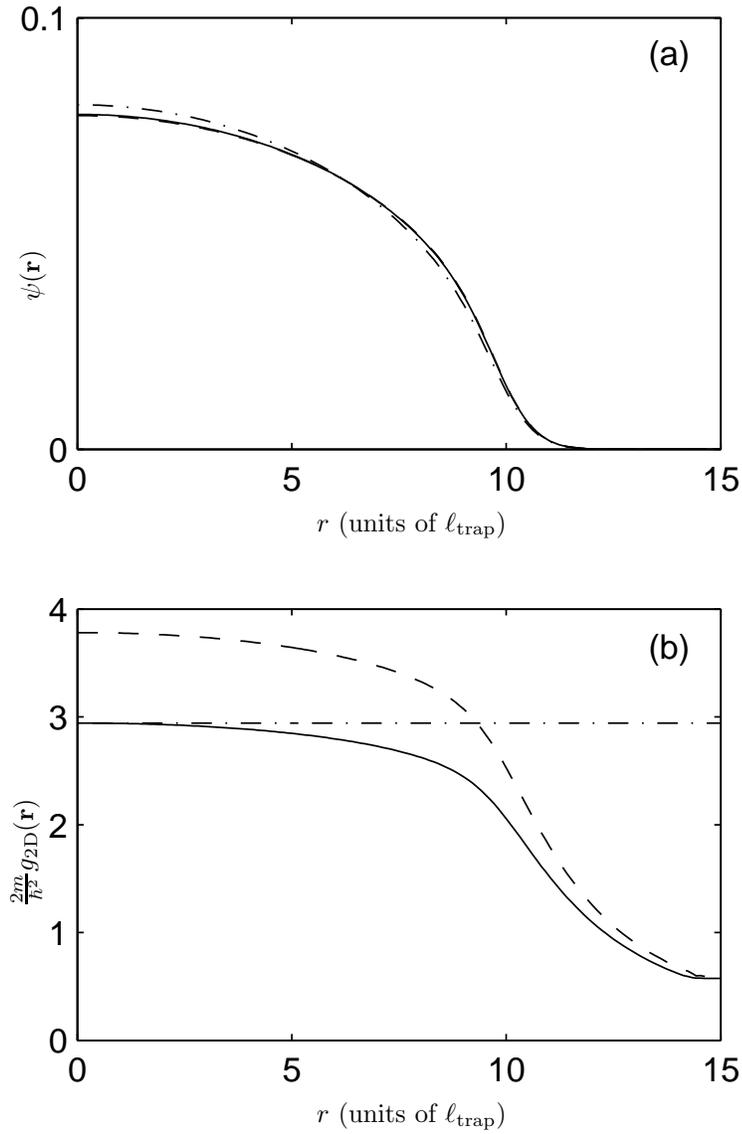,width=10cm}
\caption{(a) Wave functions $\psi(\bdr)$, and (b) coupling parameters $g(\bdr)$
calculated for $\mu=25 \hbar\omega$.  The dash-dot lines correspond to
equation~(\ref{eq:g2d}), the dashed lines correspond to
equation~(\ref{eq:gn0simple}), while the solid lines represent the full
numerical solution of equation~(\ref{eq:gn0}).  \label{fig:psiandg_mu25}}}
\end{figure}

We will solve the GPE for three different approximations for the coupling
parameter.  The simplest case restricts $g_{\rm 2D}(\bdr)$ to be constant
everywhere and determined by equation~(\ref{eq:g2d}) as in our earlier
work~\cite{lee2002}.  Spatial variations can most simply be introduced by using the
first term of the expansion in equation~(\ref{eq:gn0simple}), which corresponds to
the results of Kolomeisky \emph{et  al.\ }~\cite{kolomeisky2000,kolomeisky1992}.
Finally, the most accurate approximation is obtained using the full numerical
solution of equation~(\ref{eq:gn0}).  Figure~\ref{fig:psiandg_mu25} provides sample
solutions of the ground state wave functions and coupling parameters calculated
using these three different approximations for the same chemical potential.  It can
be seen that, although the wave functions are fairly similar, the coupling
parameters behave quite differently.  The results for the constant coupling
parameter agree well with the predictions of equation~(\ref{eq:gn0}), and differ
significantly only towards the edge of the condensate.  Of course, this is to be
expected since  $\mu \approx n_0g_{\rm 2D}(\bdr)$ near the centre of the trap.  The
energy contribution due to interactions for atoms on the edge of the condensate is
greater with the constant coupling parameter than with either of the spatially
varying parameters, and hence the constant parameter wave function has a lower
amplitude in these surface regions.

Figure~\ref{fig:psiandg_mu25} does show a large difference between the two
spatially dependent coupling parameters, however, especially in the central
region where the condensate density is greatest.  The coupling parameter of
equation~(\ref{eq:gn0simple}) is greater by about a third at the centre of the
condensate than the full expression of equation~(\ref{eq:gn0}), and remains
larger throughout.  This arises because the expansion of
equation~(\ref{eq:gn0simple}) does not converge sufficiently rapidly.  Indeed,
for the case illustrated here, the second order term in the expansion (which is
negative) reaches a magnitude of approximately $-1$ at the centre of the trap
(in the units used in figure~\ref{fig:psiandg_mu25}b).  

\begin{figure}
\center{
\epsfig{file=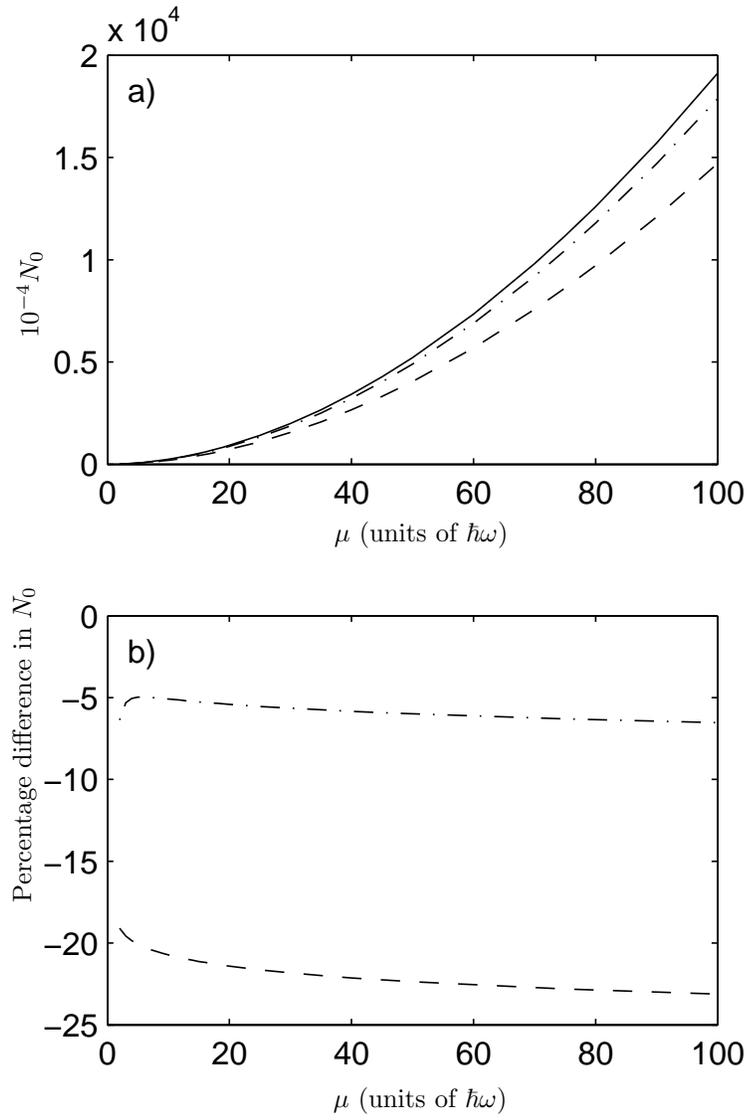,width=10cm}
\caption{a)  Condensate numbers as a function of chemical potential in 2D for
various coupling parameters.   b) The
percentage differences in the condensate populations as compared to the full
numerical solution. The line styles correspond to those used in
figure~\ref{fig:psiandg_mu25}.  \label{fig:nums}}}
\end{figure}

Such large differences in the coupling parameters can lead to problems when
calculating related quantities, such as condensate populations. 
Figure~\ref{fig:nums} shows the results for condensate populations versus chemical
potentials calculated with each of the coupling parameters.  As can be seen the
predictions obtained using the approximation of equation~(\ref{eq:gn0simple})
underestimate the condensate numbers by roughly $20$ per cent compared to the full
numerical solution of equation~(\ref{eq:gn0}).  This is a consequence of the
greater strength of the coupling parameter which occurs in this approximation. 
Agreement between equations~(\ref{eq:gn0}) and~(\ref{eq:gn0simple}) is poor even
though the usual criterion for a dilute gas ($n_0a_{\rm 2D}^2 \ll 1$) is obeyed
($n_0a_{\rm 2D}^2$ is of the order of $10^{-4}$ for the situation illustrated in
figure~\ref{fig:psiandg_mu25}).  Indeed, in order to apply the approximation in
equation~(\ref{eq:gn0simple}) we require that $\ln(n_0a_{\rm 2D}^2) \gg 1$ (while
$n_0a_{\rm 2D}^2 <1$), which is a much more stringent criterion, and one that is
experimentally unfeasible requiring at least $n_0a_{\rm 2D}^2 \lesssim 10^{-20}$. 
For this reason the use of the full expression in equation~(\ref{eq:gn0}) is
necessary to simulate real 2D gases.  

The spatially-constant approximation to the coupling parameter gives
comparatively much better agreement with the full numerical solution in
figure~\ref{fig:nums}, although it also underestimates the condensate number by
about $5$ per cent.  It would appear from these results therefore that if a
simple analytical approximation of $g_{\rm 2D}$ is required to estimate bulk
properties of the system then the spatially constant approximation is
preferable to equation~(\ref{eq:gn0simple}).  Of course the spatially constant
approximation does not deal with the boundary regions of the condensate well,
and so for properties dominated by edge effects the better analytical
approximation is likely to be that of equation~(\ref{eq:gn0simple}).

\begin{figure}
\center{
\epsfig{file=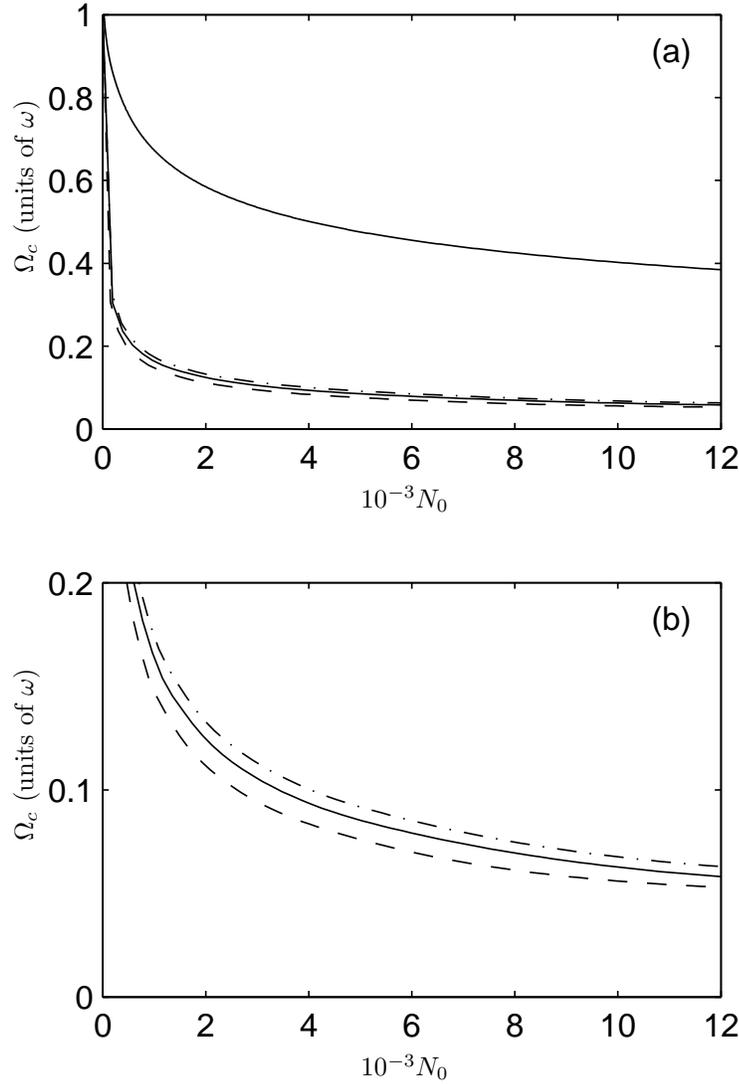,width=10cm}
\caption{a) Critical frequency versus condensate number for 2D and 3D
condensates.  The upper curve corresponds to the 3D case with $a_{\rm 3D}=
a_{\rm 2D}$, whilst the lower curves represent the various 2D results.  b) A
detailed view of the critical frequencies in 2D using the three forms of the
coupling parameter discussed.  The line styles correspond to those used in
figure~\ref{fig:psiandg_mu25}. \label{fig:criticalfreq}}}
\end{figure}

The wave functions for vortex states can also be obtained in 2D, if we assume a
solution of the form 
\begin{equation}
\psi(\bdr) = \phi(r)e^{i\kappa\theta},
\end{equation}
where $\theta$ is the angle around the vortex core, and $\kappa$ is an integer.
The phase of $\psi$ therefore wraps around by $2\pi\kappa$ as the range of
$\theta$ is traversed.  The energy per particle (in a non-rotating frame) for a
condensate with wave function $\psi$ is given by the functional
\begin{equation}
E[\psi] = \int d\bdr \left[ {\hbar^2 \over 2m}|\nabla \psi(\bdr) |^2 + V_{\rm
trap}(\bdr)|\psi(\bdr)|^2 + {N_0g_{\rm 2D}(\bdr) \over 2}|\psi(\bdr)|^4
\right].
\end{equation}
Creation of a vortex in the centre of the trap comes at the cost of increasing the
contributions from both the kinetic energy and the trapping potential terms in the
energy functional, although the interaction term is reduced by virtue of a lower
central density.  The single vortex state can be made energetically favourable by
rotating the trap at a frequency $\Omega$ such that $E[\psi_{\kappa =0}]$ becomes
less than $E[\psi_{\kappa =1}] - \Omega N_0 \hbar \kappa$.  The point at which this
occurs is known as the thermodynamic critical frequency $\Omega_c$, and this is
shown in figure~\ref{fig:criticalfreq} for the various forms of $g_{\rm 2D}$.  The
2D critical frequency is substantially lower than for a 3D gas with the same
scattering length ($a_{\rm 3D} = a_{\rm 2D}$) due principally to the much higher
interaction strength which occurs in 2D~\cite{lee2002}.  In
figure~\ref{fig:criticalfreq}b it can be seen that the effect of a
density-dependent coupling parameter is to reduce the critical frequency as
compared to the constant parameter case.  This is to be expected since the
appearance of a vortex lowers the mean density of the condensate, which decreases
the coupling parameter calculated from either of equations~(\ref{eq:gn0})
or~(\ref{eq:gn0simple}).  The lower coupling parameter means a greater saving in
the interaction energy term when a vortex is created and therefore decreases the
critical frequency.  The saving in the interaction energy is due principally to
the reduction of the density in the centre of the condensate where the density (and
hence coupling parameter) is greatest in the ground state.  Because the ground
state coupling parameter is much larger in the approximation of
equation~(\ref{eq:gn0simple}), the interaction energy saving is also greater. The
critical frequency is therefore lower in this approximation compared to the results
obtained using the coupling parameter of equation~(\ref{eq:gn0}).  In contrast to
the spatially dependent $g_{\rm 2D}$ cases, the spatially-constant coupling
parameter calculated from equation~(\ref{eq:g2d}) increases when a vortex is
formed, due to the greater chemical potential of the vortex state, and so the
critical frequency is higher in this approximation.   

\section{Conclusions}

In this paper we have applied previous results obtained for the many-body
T-matrix in a homogeneous condensate to the more currently relevant problem of a
trapped condensate, by means of a local density approximation.  This leads to a
spatially dependent coupling parameter appearing in the non-linear term of the
Gross-Pitaevskii equation.  We have shown that results obtained using the full
numerical solution to the coupling parameter can differ substantially from the
simple first approximation obtained via a series expansion.  The form of the
coupling parameter in this approximation is the same as that presented in recent
work by Kolomeisky \emph{et  al.\ }~\cite{kolomeisky2000,kolomeisky1992},
Tanatar \emph{et al.\ }~\cite{tanatar2002}, and
Shevchenko~\cite{shevchenko1992}, and is closely related to the work of
Schick~\cite{schick}.  However, this approximation is only valid in the limit
$\ln(n_0a_{\rm 2D}^2) \gg 1$ (while $n_0a_{\rm 2D}^2 <1$) which is not
experimentally relevant.  Our results indicate that the full expression of
equation~(\ref{eq:gn0}) may be needed to model current experiments accurately. 
Corrections to equation~(\ref{eq:gn0}) are expected to be of order $(na_{\rm
2D}^2)/\ln(na_{\rm 2D}^2)$, and the limit where this parameter is small should
be experimentally relevant.

Agreement with the full numerical solution for the coupling parameter is found to
be substantially better if it is approximated using a spatially constant (but
energy dependent) parameter as in the homogeneous limit.  It would seem from the
results presented here that if an approximate analytical form of the coupling
parameter is required (for deriving approximate Thomas-Fermi wave functions for
example) then the spatially constant form of $g_{\rm 2D}$ given in
equation~(\ref{eq:g2d}) is preferable to the expression of
equation~(\ref{eq:gn0simple}) for many purposes.

\ack We would like to thank K.~Burnett and M.J.~Davis for useful discussions. This
research was supported by the Engineering and Physical Sciences Research Council of
the United Kingdom, and by the European Union via the ``Cold Quantum Gases''
network. In addition, M.D.L.\ was supported by the Long Studentship from The
Queen's College, Oxford, and S.A.M.\ would like to thank the Royal Society for
financial support.

\end{document}